\documentclass[sigconf]{acmart}

\usepackage{booktabs} 

\usepackage{algorithm}

\usepackage{algorithmic}

 \usepackage{mathrsfs}

\usepackage{amsfonts,amssymb}

\usepackage{amsmath,bm}
\usepackage{latexsym}

\usepackage{subfigure}

\usepackage[english]{babel}
\usepackage[colorinlistoftodos]{todonotes}
\usepackage{multirow}

\AtBeginDocument{%
  \providecommand\BibTeX{{%
    \normalfont B\kern-0.5em{\scshape i\kern-0.25em b}\kern-0.8em\TeX}}}

\copyrightyear{2020}
\acmYear{2020}
\setcopyright{acmcopyright}\acmConference[KDD '20]{Proceedings of the 26th ACM SIGKDD Conference on Knowledge Discovery and Data Mining}{August 23--27, 2020}{Virtual Event, CA, USA}
\acmPrice{15.00}
\acmDOI{10.1145/3394486.3403329}
\acmISBN{978-1-4503-7998-4/20/08}

\begin{document}

\title[Maximizing Cumulative User Engagement]{Maximizing Cumulative User Engagement in Sequential Recommendation: An Online Optimization Perspective}

\author{Yifei Zhao}
\affiliation{%
  \institution{MIND, Alibaba Group}
  andy.zyf@alibaba-inc.com
}
\author{Yu-Hang Zhou}
\affiliation{%
  \institution{MIND, Alibaba Group}
  zyh174606@alibaba-inc.com
}
\author{Mingdong Ou}
\affiliation{%
  \institution{MIND, Alibaba Group}
  mingdong.omd@alibaba-inc.com
}
\author{Huan Xu}
\affiliation{%
  \institution{MIND, Alibaba Group}
  huan.xu@alibaba-inc.com
}
\author{Nan Li}
\affiliation{%
  \institution{STCA, Microsoft}
  alex.nan@microsoft.com
}

\renewcommand{\shortauthors}{Zhao and Zhou, et al.}

\begin{abstract}
To maximize cumulative user engagement (e.g. cumulative clicks) in sequential recommendation, it is often needed to tradeoff two potentially conflicting objectives, that is, pursuing higher immediate user engagement (e.g., click-through rate) and encouraging user browsing (i.e., more items exposured). Existing works often study these two tasks separately, thus tend to result in sub-optimal results.
In this paper, we study this problem from an online optimization perspective, and propose a flexible and practical framework to explicitly tradeoff longer user browsing length and high immediate user engagement. Specifically, by considering items as actions,  user's requests as states and user leaving as an absorbing state, we formulate each user's behavior as a personalized {\it Markov decision process} (MDP), and the problem of maximizing cumulative user engagement is reduced to a {\it stochastic shortest path} (SSP) problem. Meanwhile,  with immediate user engagement and quit probability estimation, it is shown that the SSP problem can be efficiently solved via dynamic programming.  Experiments on real-world datasets demonstrate the effectiveness of the proposed approach. Moreover, this approach is deployed at a large E-commerce platform,  achieved over $7\%$ improvement of cumulative clicks.
\end{abstract}




\maketitle

\section{Introduction}
In recent years, sequential recommendation has drawn attention due to its wide application in various domains~\cite{chen2018sequential,tang2018personalized,huang2018improving,donkers2017sequential,ebesu2018collaborative}, such as E-commerce, social media, digital entertainment and so on. It even props up popular stand-alone products like \textit{Toutiao}\footnote{\url{https://www.toutiao.com/}}, \textit{Tik Tok}\footnote{\url{https://www.tiktok.com/}} and so on.


\begin{figure}
\setlength{\abovecaptionskip}{0cm}   
\setlength{\belowcaptionskip}{-0.5cm}   
\centering
\includegraphics[width=0.45\textwidth]{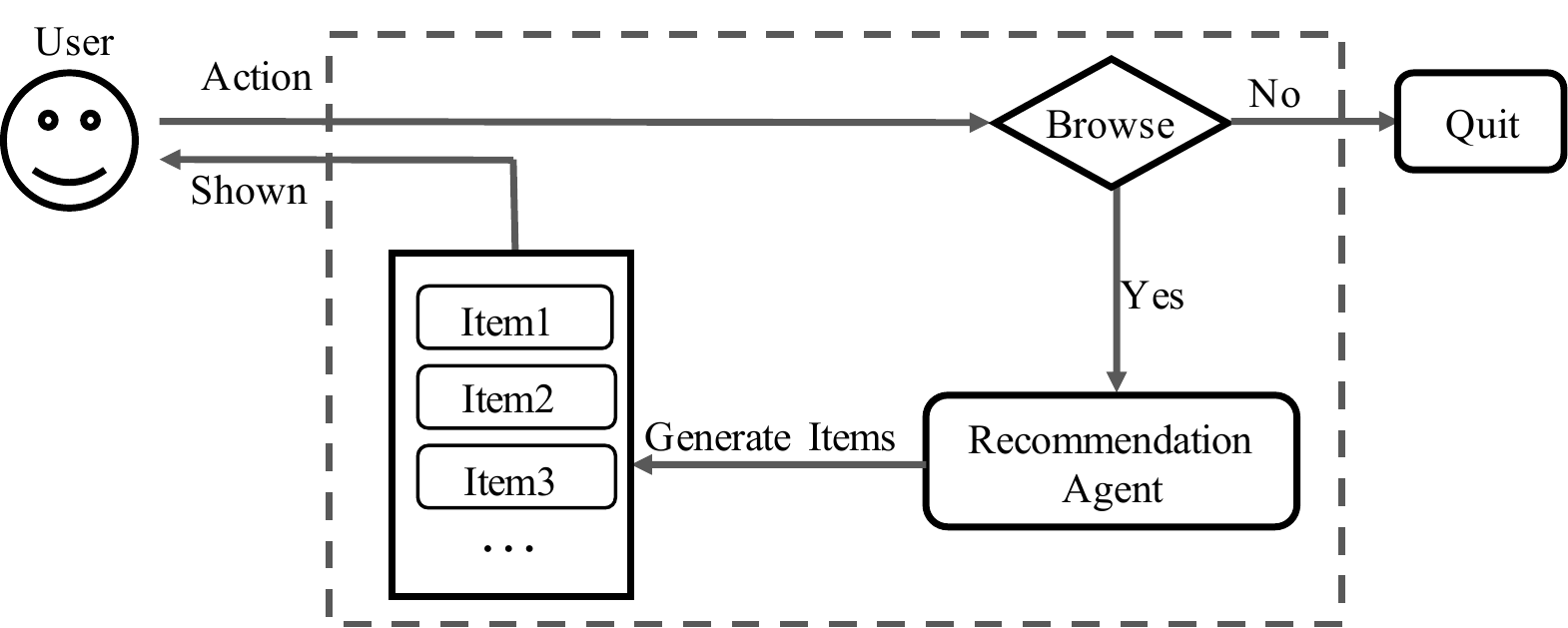}
\caption{Sequential Recommendation Procedure}\label{fig:sr}
\end{figure}



Different from traditional recommender systems where it is often assumed that the number of recommended items is fixed,  the most important feature of sequential recommendation is that it iteratively recommend items until the user quits (as depicted in  Figure~{\ref{fig:sr}}), which means that users can browse endless items if they want.
Its goal is to maximize cumulative user engagement in each session, such as cumulative clicks, cumulative dwell time, etc. To this end, the recommender systems need to simultaneously achieve two objectives:
\begin{enumerate}
	\item[a)] Attracting users to have a longer session such that more items can be browsed;
	\item[b)] Capturing user interests such that higher immediate user engagement can be achieved.
\end{enumerate}
%

In traditional recommender systems, since the number of recommended items is fixed, most efforts are spent on improving immediate user engagement, which is often measured by click-through rate, etc.
However, when such a strategy is adopted for sequential recommendation, it tends to result in sub-optimal cumulative user engagement, due to the limited number of browsed items. Moreover, due to their inherent conflicts, it is not a trivial task to achieve a longer session and higher immediate user engagement simultaneously (which can be demonstrated in the experiments).
%
%
For example, to achieve a longer session, it is generally needed to explore diverse recommendation results; almost for sure, this will sacrifice immediate user engagement.
Therefore, how to tradeoff a longer session and higher immediate user engagement becomes critical to achieve higher cumulative user engagement, and this is essentially the key problem of sequential recommendation.


Generally speaking, existing works on sequential recommendation fall into two groups. The first group of works try to leverage sequential information (e.g.,  users' interaction behavior) to estimate the probability of user engagement (e.g., click-through rate) more accurately ~\cite{chen2018sequential,tang2018personalized,huang2018improving,donkers2017sequential,ebesu2018collaborative}, for example, by using recurrent neural network or its variants~\cite{chen2018sequential,wu2017recurrent,donkers2017sequential}. By exploiting the sequential behavior pattern, these methods focus on capturing user interests more accurately, but do not consider to extend session length thus may lead to sub-optimal results. Based on the observation that diverse results tend to attract users to browse more items, the second group of methods explicitly considers the diversity of recommendation results~\cite{teo2016adaptive,devooght2017long,carbonell1998use}.  However, the relationship between diversity and user browsing length is mostly empirical; thus it is not so well-founded to optimize diversity directly, especially when it is still a fact that there are no well-accepted diversity measures so far. Therefore, it is still a challenge to optimize cumulative user engagement in the sequential recommendation.


In this paper, we consider the problem of maximizing cumulative user engagement in sequential recommendation from an online optimization perspective, and propose a flexible and practical framework to solve it. Specifically, by considering  different items as different actions,  user's different requests as states and user leaving as an absorbing state, we consider user browsing process in the {\it Markov decision process} (MDP) framework, and as a consequence the problem of maximizing cumulative user engagement can be reduced to a {\it stochastic shortest path} (SSP) problem. To make this framework practical,  at each state (except absorbing state), we need to know two probabilities for each possible action, i.e., the probability of achieving user engagement (e.g., click) and the probability of transitioning to the absorbing state which means that the user quits the browsing process. Obviously, the problem of estimating the probability of user engagement has been well studied, and many existing machine learning methods can be employed. Meanwhile, we propose a multi-instance learning method to estimate the probability of transitioning to the absorbing state (i.e., user quit model). With this framework and corresponding probabilities effectively estimated, the SSP problem can be solved efficiently via dynamic programming.  Experiments on real-world datasets and an online E-commerce platform demonstrate the effectiveness of the proposed approach.

In summary, our main contributions are listed as below:
\begin{itemize}
\item We solve the problem of maximizing cumulative user engagement within an online optimization framework. Within this framework, we explicitly tradeoff longer user browsing session and high immediate user engagement to maximize cumulative user engagement in sequential recommendation.
\item Within the online optimization framework, we propose a practical approach which is efficient and easy to implement in real-world applications.  In this approach,  existing works on user engagement estimation can be exploited,  a new multi-instance learning method is used for user quit model estimation, and the corresponding optimization problem can be efficiently solved via dynamic programming.
\item Experiments on real-world dataset demonstrate the effectiveness of the proposed approach,  and detailed analysis shows the correlation between user browsing and immediate user engagement. Moreover, the proposed approach has been deployed on a large E-commerce platform, and achieve over $7\%$ improvement on cumulative clicks.
\end{itemize}

The rest of the paper is organized as follows. In Section 2, we discuss some related works. Problem statement is given in Section 3. Section 4 provides the framework named MDP-SSP and the related algorithms. Experiments are carried out in Section 5, and finally we give a conclusion in Section 6.

\section{Related work}
\subsection{Sequential Recommendation}

In recent years, conventional recommendation methods, e.g., 
RNN models~\cite{hidasi2015session,donkers2017sequential,huang2018improving}, memory networks with attention~\cite{chen2018sequential,ebesu2018collaborative,huang2018improving}, etc., are applied in sequential recommendation scenarios widely.
In order to find the \textit{next item} that should be recommended, 
RNN models capture the user's sequence pattens by utilizing historic sequential information. One could also train a memory network and introduce the attention mechanism to weighting some sequential elements.
~\cite{tang2018personalized,donkers2017sequential} show that these methods significantly outperform the classic ones which ignored the sequential information.
Essentially, they are still estimating the immediate user engagement (i.e. click-through rate) on the \textit{next item}, without considering quit probability. Therefore further improvements are necessary to maximize cumulative user engagement.

\subsection{MDP and SSP}
Stochastic Shortest Path (SSP) is a stochastic version of the classical shortest path problem: for each node of a graph, we must choose a probability distribution over the set of successor nodes so as to reach a certain destination node with minimum expected cost~\cite{bertsekas1991analysis,polychronopoulos1996stochastic}. SSP problem is essentially a Markov Decision Process (MDP) problem, with an assumption that there is an absorbing state and a proper strategy. Some variants of Dynamic Programming can be adopted to solve the problem~\cite{bonet2003labeled,kolobov2011heuristic,trevizan2016heuristic,barto1995learning}.
Real Time Dynamic Program (RTDP) is an algorithm for solving non-deterministic planning problems with full observability, which can be understood either as an heuristic search or as a dynamic programming (DP) procedure~\cite{barto1995learning}. Labeled RTDP~\cite{bonet2003labeled} is a variant of RTDP, and the key idea is to label a state as \textit{solved} if the state and its successors have converged, and the solved states will not be updated further.

\subsection{Multi-Instance Learning}
In Multi-instance learning (MIL) tasks, each example is represented by a bag of instances~\cite{dietterich1997solving}. A bag is positive if it contains at least one positive instance, and negative otherwise. The approaches for MIL can fall into three paradigms according to~\cite{amores2013multiple}: the instance-space paradigm
, the bag-space paradigm
and the embedded-space paradigm.
For our sequential recommendation setting, the need of modeling the transiting probability is in accordance with the instance-space paradigm. Several SVM based methods are proposed in instance-level MIL tasks~\cite{maron1998framework,zhang2002dd,andrews2003support,bunescu2007multiple}. 
MI-SVM is a variant of SVM-like MIL approaches, the main idea is that it forces the instance farthest to the decision hyperplane (with the largest margin) to be positive in each iteration.

\section{Problem Statement}
We model each browsing process as a personalized {\it Markov Decision Process} (MDP) including an absorbing state, and consider the problem of maximizing cumulative user engagement as a {\it stochastic shortest path} (SSP) problem.

\subsection{Personalized MDP Model}
The MDP consists of a tuple with four elements (S, A, R, P):
\begin{itemize}
\item \textbf{State space} $\mathcal{S}: \mathcal{S}=\{s_1, s_2 , s_3,..., s_t,..., s_T, s_A\}$. Here we take each step in the recommendation sequence as an individual state and define $s_t=t$, where $t$ is the step index. Since only one item is shown to the user in each step, $t$ is also the sequence number of the browsed items. $T$ is the upper limit of the browsing session length, which is large enough for the recommendation senarioes. $s_A$ is defined as the absorbing state meaning that the user has left.

\item \textbf{Action space} $\mathcal{A}: \mathcal{A}=\{1,2,3,...,K\}$. Action space $\mathcal{A}$ contains all candidates that could be recommended in the present session.

\item \textbf{Reward} $\mathcal{R}: \mathcal{R}\in\mathbb{R}^{(T+1)\times K}$. Denote $s$ as a state in $\mathcal{S}$, and $a$ as an action in $\mathcal{A}$, and then $R_{s, a}$ is the reward after taking action $a$ in state $s$. Specifically, $R_{s_t, a_t}$ is the immediate user engagement (e.g., click-through rate) in the $t$-th step.

\item \textbf{Transition probability} $\mathcal{P}: \mathcal{P}\in\mathbb{R}^{(T+1)\times K\times(T+1)}$, and $P_{s, a, s'}\in [0,1]$ is the probability of transiting from state $s$ to state $s'$ after taking action $a$.
\end{itemize}

Since the states in $\mathcal{S}$ are sequential, we introduce a regulation on $\mathcal{P}$ that from all states except $s_T$ and $s_A$, users can only transit to the next state (go on browsing) or jump into the absorbing state (quit). Moreover, from the last browsing step, users could only be admitted to jump into absorbing state. Formally, we have
\vspace{-0.1cm}
\begin{equation}
  \label{eqn:quit}
  \left\{
  \begin{aligned}
    P_{s_i, a_i, s_{i+1}}+P_{s_i, a_i, s_A}=1,\ i<T \\
    P_{s_T, a_t, s_A}=1,\ i=T
  \end{aligned}
  \right.
\end{equation}

\begin{figure}
\setlength{\abovecaptionskip}{0cm}   
\setlength{\belowcaptionskip}{-0.5cm}   
\includegraphics[width=0.5\textwidth]{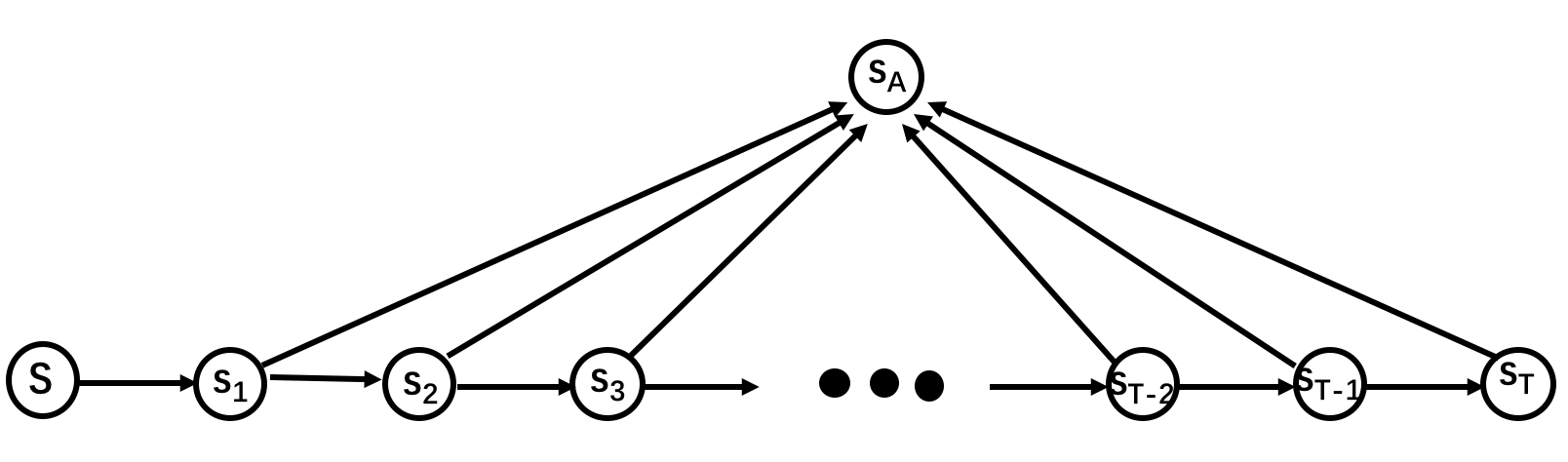}
\caption{MDP Finite-State Machine}\label{fig:SSP}
\end{figure}

The finite-state machine of the procedure is shown as Figure~{\ref{fig:SSP}}. Furthermore, it should be emphasized that the proposed MDP model is personalized and we will infer a new MDP model for each online session. An efficient algorithm for generating MDP models will be presented later.

\subsection{SSP Problem}
Based on the MDP model, the optimization of cumulative rewards in a sequential recommendation can be formally formulated as a SSP problem:
Given a MDP, the objective is to find a policy $\pi^*$: $\mathcal{S} \rightarrow \mathcal{A}$, which can help us to plan a path with maximal cumulative rewards, i.e.
\vspace{-0.1cm}
\begin{equation}
\label{eqn:IPVexpectation}
\pi^* = \arg\max_{\pi}\mathbb{E}(\sum_{t=1}^{\tau}{R_{s_t, a_t}})\ ,
\end{equation}
where $\tau$ is the actual browse length. The distribution of $\tau$ can be derived as
\vspace{-0.1cm}
\begin{align}
\mathbb{P}(\tau\ge t) = \prod_{i<t}P_{s_i, a_i, s_{i+1}}\ .
\end{align}
Thus the expected cumulative rewards in Equation~(\ref{eqn:IPVexpectation}) can be represented as
\vspace{-0.1cm}
\begin{equation}
\label{eqn:ct}
\mathbb{E}(\sum_{t=1}^{\tau}{R_{s_t, a_t}}) = \sum_{t\le T}{R_{s_t, a_t}\mathbb{P}(\tau\ge t)}\ ,
\end{equation}
Finally, by introducing Equation~(\ref{eqn:quit}) into Equation~(\ref{eqn:ct}), we have
\vspace{-0.1cm}
\begin{equation}
\label{eqn:IPV2}
 \mathbb{E}(\sum_{t=1}^{\tau}{R_{s_t, a_t}}) = \sum_{t=1}^{T} R_{s_t, a_t} \times \prod_{i<t}(1-P_{s_i, a_i, s_A}) \ .
\end{equation}
\vspace{-0.4cm}

\subsubsection{Remark 1.} Maximize Equation~(\ref{eqn:IPV2}) is simultaneously optimizing two points mentioned in \textbf{Introduction}:
1) user browse length, i.e.  $\tau$, and 2) immediate user engagement, i.e. $R_{s_t, a_t}$.

According to the formulation, we should first estimate $R_{s_t, a_t}$ and $P_{s_i, a_i, s_A}$ in Equation~(\ref{eqn:IPV2}), which is essentially generating a personalized MDP model. Then we optimize a policy by maximizing Equation~(\ref{eqn:IPV2}), which could be used to plan a recommendation sequence $[a_1,\cdots,a_T]$ (or called {\it Path} in SSP) to the corresponding user.

\section{The Proposed Approach} \label{algSec}

In this section, we first propose an online optimization framework named {\it MDP-SSP} considering browsing session length and immediate user engagement simultaneously and maximizing the cumulative user engagement directly. Then the related algorithms are presented detailedly.

\subsection{MDP-SSP Framework}

\begin{figure}
\setlength{\abovecaptionskip}{0cm}   
\setlength{\belowcaptionskip}{-0.5cm}   
\centering
\includegraphics[width=0.4\textwidth]{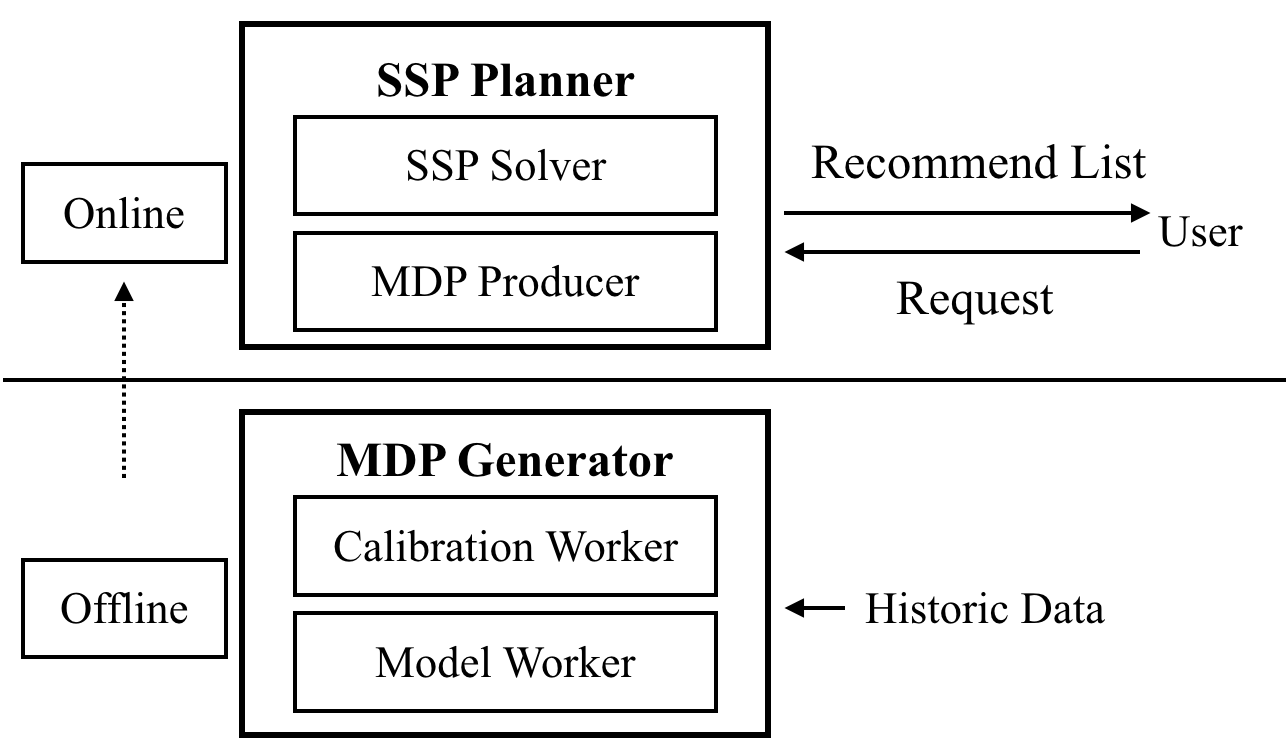}
\caption{MDP-SSP Framework}
\label{fig:framework3}
\end{figure}

In order to maximize the expected cumulative rewards, as mentioned previously, we should learn a MDP generator from which the personalized MDP model can be generated online, and then plan the recommendation sequence with the personalized MDP model. Therefore, the proposed MDP-SSP framework consists of two parts: an offline MDP Generator and an online SSP Planner, which is shown in Figure~{\ref{fig:framework3}}.

\subsubsection{MDP Generator}
is designed to generate personalized MDPs for each online sessions. There are two submodules in this part: \textit{Model Worker} and \textit{Calibration Worker}. \textit{Model Worker} is used to learn a model from offline historic data, aiming to provide necessary elements of the personalized MDP. Specifically, the reward function $R_{s_t, a_t}$ and the quit probability $P_{s_i, a_i,s_A}$ in Equation~(\ref{eqn:IPV2}) are needed. Here $R_{s_t, a_t}$ could be an immediate user engagement, e.g. immediate click, thus \textit{Model Worker} contains the corresponding estimation model, e.g. click model. In the meanwhile, $P_{s_i, a_i,s_A}$ is related to a quit model which determines the browse session length and is an important component of \textit{Model Worker}.

Moreover, since the efficiency of SSP planning depends on the accuracy of the generated MDP model, we introduce an additional \textit{Calibration Worker} to calibrate the ranking scores obtained from the learned model to the real value~\cite{niculescu2005predicting,lin2007note,he2014practical}.

\subsubsection{SSP Planner} plans a \textit{shortest path} (with maximal cumulative rewards) consisting of sequential recommended items. It also contains two submodules: \textit{MDP Producer} and \textit{SSP Solver}. Based on the generator learned by the offline MDP Generator algorithm, \textit{MDP Producer} generates a personalized MDP for the user of present session. Then \textit{SSP Solver} will plan an optimal path based on the personalized MDP to the user.

\subsection{Offline MDP Generator Algorithm}

In this section, we present an offline algorithm to learn the reward function $R_{s_t, a_t}$ and the quit probability $P_{s_i, a_i, s_A}$, which are required to generate online personalized MDPs. We will see that the problem of modeling $P_{s_i, a_i, s_A}$ is more critical and difficult. In practice, the historic data we obtain is often an $item\ set$ containing the items seen by the user until the end of a session, which would make the users quit or go on browsing. However, it is hard to know which item in the $item\ set$ is exactly the chief cause. In order to estimate the quit probability for each item, we alternatively adopt the Multi-Instance Learning (MIL) framework by taking $item\ set$ as $bag$ and $item$ as $instance$. Detailedly, if the $item\ set$ causes a quit, the user dislikes all the items in this set; if the $item\ set$ causes a continues browse, at least one item in the $item\ set$ is accepted by the user, which is consistent with the MIL setting.

\subsubsection{Remark 2.} The standard MIL assumption states that all negative bags contain only negative instances, and that positive bags contain at least one positive instance.

By utilizing some classical MIL techniques, we can obtain the following user quit model.

\subsubsection{User Quit Model}

Based on the users' browse history, we can get sequences consisted with bags $B_i$, and one may verify that only the last bag in a browse session cannot keep the users going on browsing. We assume that the bag which can keep the user is a positive bag, written as $B_i^+$, and the last one is the negative bag written as $B_{leave}^-$, so a browse session is $B=(B_1^+, ...,B_i^+,...,B_{leave}^-)$. Our job is to construct a model to predict the quit probability for each new instance $B_{*,j}$. However, there exists a gap that the training labels we have are in bag level, while the predictions we need are in instance level. In order to cope with this problem, we introduce MI-SVM~\cite{andrews2003support} to help us train an instance level model with the bag level data, which is a novel application of MIL to recommendation to the best of our knowledge. The process for quit model training  is shown in \textbf{Algorithm~{\ref{alg:Training Data for User Quit Model}}}.


\begin{algorithm}[t]
\caption{User Quit Model}
\label{alg:Training Data for User Quit Model}
\begin{algorithmic}[1]
\REQUIRE ~~\\ 
 {Historic browse session set: $\mathcal{H}$=\{($B_1^+$, $B_2^+$, ...,$B_i^+$,...,$B_{leave}^-$)\}}
\STATE $\mathcal{H}$ is converted to $\mathcal{H}_0$ by NSK~\cite{gartner2002multi}, and a initial $SVM_{0}$ is obtained
\STATE $t=1$
\FORALL {$B_{\_t} \in \mathcal{H}$}
\FORALL{$B_i \in B_{\_t}$}
\STATE Select $B_{i,j}^+$ with maximum value according to $SVM_{t-1}$
\STATE $B_{i_{t, max}}^+ = B_{i,j}^+$
\ENDFOR\\
\STATE $B_{\_t}=( \cdots,B_{i_{t, max}}^+,\cdots ,B_{leave}^-)$
\ENDFOR\\
\STATE $\mathcal{H}_t = \{B_{\_t}\}$
\STATE Train $SVM_{t}$ based on $\mathcal{H}_t$
\STATE $t = t+1$
\REPEAT
\STATE line 3-12
\UNTIL ${\mathcal{H}_t =\mathcal{H}_{t-1} }$
\RETURN$SVM_{t}$
\end{algorithmic}
\end{algorithm}

\subsubsection{Model Calibration}
In the industrial recommendation system, ranking scores provided by the click model and quit model are not equivalent to the reward $R_{s_t, a_t}$ and the transition probability $P_{s_i, a_i, s_A}$ of MDPs. Thus it is necessary to calibrate the model outputs to real probabilities. Readers interested in this topic may go to~\cite{niculescu2005predicting,lin2007note,he2014practical} for details. In this paper, denoting the predicted score as \textit{$f(B_{i,j})$}, the real probability value can be represented as follow:
\vspace{-0.1cm}
 \begin{equation}
  \textit P(\textit y=1|B_{i,j}) = \frac{1}{1+exp(\textit A*f(B_{i,j})+\textit B)} \ ,
\end{equation}
where \textit A and \textit B are two scalar parameters can be learned from historic data.

\subsection{Online SSP Planner Algorithm}
Based on the MDP Generator discussed in the last subsection, we formally introduce SSP Planner, which consists of MDP Producer and SSP Solver.

\subsubsection{MDP Producer}
When a new session comes, the MDP Producer receives online information of user and items from server, and then feeds them into the generators derived from MDP Generator. After that, the reward and transition probability can be obtained and a personalized MDP is produced in real time. It's worth noting that, the information about how many items the user has browsed, how many times the item's category have been shown to the user and clicked by the user, should be considered. These interactive features play an important role in causing the user go on browse or quit intuitively.

\subsubsection{SSP Solver}
From MDP Producer we can get a personalized MDP for the present session, and the next job is to find a path $[a_1,\cdots,a_T]$ with maximal cumulative rewards. Except absorbing state, the corresponding MDP has T states, and then optimal state value function can be addressed with dynamic programing in $T$-steps interaction. Furthermore, it is easy to verify that the transition matrix of our specifically designed MDP preserves an upper triangular structure, shown as Equation~({\ref{eqn:upper matrix}}).
\vspace{-0.1cm}
\begin{equation}       
\left[  \begin{array}{cccccc}   
    0 & P_{s_1, s_2} & 0 & 0 &  \cdots & P_{s_1, s_A}\\  
    0 & 0 & P_{s_2, s_3} & 0 &  \cdots & P_{s_2, s_A}\\  
    \cdots & \cdots & \cdots & P_{s_i, s_{i+1}} &  \cdots & P_{s_i, s_A}\\  
    \cdots & \cdots & \cdots & \cdots &  \cdots & P_{s_T, s_A}\\  
    \cdots & \cdots & \cdots & \cdots &  \cdots & \cdots  
  \end{array}
\right]           
\label{eqn:upper matrix}
\end{equation}
\vspace{-0.1cm}

Based on the special structured transition matrix, it is easy to find that the latter state value function will not change when we update the current state value function. Therefore the backwards induction could be adopted. One may start from the absorbing state, and iteratively obtain the optimal policy as well as the related optimal state value function. We formally summarize this procedures as follow:
\vspace{-0.1cm}
\begin{equation}
\label{eqn:VA}
V^*(s_A) = 0\ .
\end{equation}
Further more, when $i=T$, we have
\vspace{-0.1cm}
\begin{align}
\pi^*(s_T) & =\arg\max_{a_T}\ \{R_{s_T, a_T}+P_{s_T, a_T, s_A} V^*(s_A)\}\notag  \\
	& =\arg\max_{a_T}\ \{R_{s_T,a_T}\} \label{eqn:best policy1}\ ,\\
V^*(s_T)& =\max_{a_T}\ \{R_{s_T,a_T}\}\label{eqn:best value1}\ ,
\end{align}

\vspace{-0.3cm}
\leftline {when $i<T$, we have}
\vspace{-0.5cm}
\begin{align}
\pi^*(s_t) & =\arg\max_{a_t}\ \{R_{s_t, a_t}+P_{s_t, a_t, s_{t+1}} V^*(s_{t+1})\notag \\
	&\quad   +P_{s_t, a_t, s_A} V^*(s_A)\}\notag  \\
	& =\arg\max_{a_t}\ \{R_{s_t, a_t} + P_{s_t, a_t, s_{t+1}} V^*(s_{t+1})\}\label{eqn:best policy}\ ,\\
V^*(s_t) & =\max_{a_t}\ \{R_{s_t, a_t}+P_{s_t, a_t, s_{t+1}} V^*(s_{t+1})\}\label{eqn:best value}\ .
\end{align}
\vspace{-0.3cm}

Based on the Equations~{(\ref{eqn:VA})}-{(\ref{eqn:best value})}, we can plan an optimal path  $[a_1,\cdots,a_T]$. The optimization procedure is shown in \textbf{Algorithm}~{\ref{alg:SSP Solver}}. We can see that the whole planning procedure is quite simple and clear, which benefits the online application of the proposed method.
Specifically, assuming there are $K$ candidates, the complexity of SSP is $O(TK)$ .

\begin{algorithm}[t]
\caption{SSP Solver}
\label{alg:SSP Solver}
\begin{algorithmic}[1]
\REQUIRE ~~\\ 
{User Request, MDP Generator}
\STATE Generating a personalized MDP for the current user
\STATE Initialize a vector \textit {Path} with length T
\STATE Obtain an optimal policy, {\it i.e.}, $\pi^*(s_T)=a_T$ according to Equation~({\ref{eqn:best policy1}})
\STATE Obtain an optimal state value $V^*(s_T)$ according to Equation~({\ref{eqn:best value1}})
\STATE Update $\textit{Path}[T]=a_T$
\FOR{\textit {t=T-1, ..., 2, 1}}
\STATE Obtain an optimal policy, {\it i.e.},  $\pi^*(s_t)=a_t$ according to Equation~({\ref{eqn:best policy}})
\STATE Obtain an optimal state value $V^*(s_t)$ according to Equation~({\ref{eqn:best value}})
\STATE Update $\textit{Path}[t]=a_t$
\ENDFOR
\RETURN \textit {Path} = $[a_1,\cdots,a_T]$
\end{algorithmic}
\end{algorithm}

\section{Experiments}
The experiments are conducted on a large E-commerce platform. We first analyze the characteristics of data which demonstrates the necessity of applying SSP,  and then evaluate SSP offline and online. 
\vspace{-0.3cm}
\subsection{Data Set}
\textbf{Dataset 1}: This dataset is for MDP Generator. It consists of 15 days historic data of user item interactions, based on which we may learn models for predicting the click-through rate and quit probability of any user item pair. \\
\textbf{Dataset 2}: This dataset is for SSP offline evaluation. We collect the active users and their corresponding browse sessions, and discard those that are inactive or excessive active. The sampling is according to the criterion that whether the browse session length is between 50 items and 100 items. Finally, we get 1000 users and corresponding browse sessions. The average length of the browse sessions is 57.\\
\textbf{Dataset 3}: This dataset is for SSP online evaluation. It is actually online environment, and has about ten millions of users and a hundred millions of items each day.

Various strategies (including SSP) will be deployed to rerank the personalized candidate items for each user in Dataset 2 and Dataset 3, to validate their effect on maximizing cumulative user engagement. Before that we should first verify the datasets are in accordance with the following characteristics: \\
\vspace{-0.3cm}
\begin {itemize}
\item Discrimination: Different items should provide different quit probabilities, and they should have a significant \textit{discrimination}. Otherwise quit probability is not necessary to be considered when make recommendations.
\item Weakly related: The quit probability of an item for a user should be \textit{weakly related} with click-through rate. Otherwise SSP and Greedy will be the same.
\end {itemize}

\subsection{Evaluation Measures}
In the experiment, we consider the cumulative clicks as cumulative user engagement. Moreover, we name cumulative clicks as $IPV$ for short, which means {\it Item Page View} and is commonly used in industry. Browse length($BL$, for short) is also a measurement since $IPV$ can be maximized through making users browse more items.

In offline evaluation, assuming that the recommended sequence length is $T$, with Equation~(\ref{eqn:quit})-(\ref{eqn:IPV2}) we have
\vspace{-0.1cm}
\begin{align}
IPV  & = \sum_{t=1}^{T} R_{s_t, a_t} \times \prod_{i<t}(1-P_{s_i, a_i, s_A}) \label{eqn:IPV3} \ ,\\
BL & =  \sum_{t=1}^{T} \prod_{i<t}(1-P_{s_i, a_i, A})\ .
\end{align}
Furthermore, define the $CTR$ of the recommended sequence as
\vspace{-0.1cm}
\begin{align}
CTR=\frac{IPV}{BL}\label{eqn:ctr}\ .
\end{align}
\vspace{-0.1cm}

In online evaluation, $IPV$ can be counted according to the actual online situation, follows that
\vspace{-0.1cm}
\begin{align}
IPV=\sum_{t=1}^{\tau}{c_t}\ ,
\end{align}
where $c_t \in \{0, 1\}$ indicates the click behavior in step t, and $\tau$ is the browse length, i.e. $BL=\tau$.

\subsection{Compared Strategies}
\begin {itemize}
\item Greedy: The key difference between our methods and traditional methods is that: we take user's quit probability into consideration and plan a path by directly maximizing $IPV$, while most other methods try hard to estimate each step reward $R_{s_t, a_t}$ as exact as possible. However when planning they just rank the items greedily according to $R_{s_t, a_t}$, ignoring that $P_{s_i, a_i, s_A}$ is also crucial to $IPV$. Greedy is the first compared strategy in which quit probability $P_{s_i, a_i, s_A}$ cannot be involved. Assuming that there are $K$ candidates and the length of planning path is $T$, the complexity is $O(TK)$.
\item Beam Search:  It is a search algorithm that balances performance and consumption. Its purpose is to decode relatively optimal paths in sequences. It is chosen as the compared strategy because the quit probability $P_{s_i, a_i, s_A}$ can be involved. We calculate beam path score according to Equation~{(\ref{eqn:IPV3})}, so that Beam Search applied here directly optimize $IPV$. Assuming that there are $K$ candidates and the length of planning path is $T$, the complexity is $O(STK)$, where $S$ is beam size.
\end{itemize}

\subsection{MDP Generator Learning}
%

We first describe MDP Generator learning in $section\ 5.4.1$ and $section 5.4.2$, with which we verify the characteristics of datasets in  $section\ 5.4.3$ and $section 5.4.4$.

\subsubsection{Model Learning}
In model learning, we take full use of user attributes and item attributes. Further more, we add interactive features, for example how many times the item's category have been shown to the user and clicked by the user, which intuitively play a important role in making the user go on browse. Area Under the Curve (AUC)\footnote{https://en.wikipedia.org/wiki/Receiver\_operating\_characteristic\#Area\_under\_the\_curve}, which is a frequently-used metric in industry and research, is adopted to measure the learned model, and the result is shown in \textbf{Table~{\ref{tab:CTR Model and Browse Model}}}.

\begin{table}
  \caption{CTR Model and Browse Model}
  \label{tab:CTR Model and Browse Model}
  \begin{tabular}{cl}
    \toprule
    Model & AUC\\
    \midrule
    CTR Model & 0.7194\\
    Quit Model & 0.8376\\
  \bottomrule
\end{tabular}
\end{table}

Here we state briefly about {\it Quit Model} testing method. As we indeed do not know which item makes the user go on browsing in practice, thus AUC cannot be directly  calculated on instance level. It is more rational to calculate AUC in bag level with instance prediction, as we can assume that \textit{the bag is positive if it contains at least one positive instance, the bag is negative if all the instances are negative}.

Furthermore, we take a comparison experiment on {\it Quit Model} to show the necessary of adopting MIL. As bag labels are known, the most intuitive idea is that using the bag's label to represent the instance's label. Based on this idea, we obtain a $Quit Model_{no\_MIL}$, and AUC is calculated also in bag level. The results are shown in \textbf{Table~{\ref{tab:Quit Model MIL}}}, from which we can see adopting MIL gives a improvement for {\it Quit Model} learning. 

\begin{table}
  \caption{Quit Model Comparison}
  \label{tab:Quit Model MIL}
  \begin{tabular}{cl}
    \toprule
    Model & AUC\\
    \midrule
    $Quit Model_{no\_MIL}$ & 0.8259\\
    Quit Model & 0.8376\\
  \bottomrule
\end{tabular}
\end{table}

\subsubsection{Model Calibration}

Calibration tries to map the ranking scores obtained from models to real value. It is very important here for the error will accumulate, see Equation~({\ref{eqn:best policy}})\~{}({\ref{eqn:best value}}). We apply platt scaling method, and Root Mean Square Error (RMSE)\footnote{https://en.wikipedia.org/wiki/Root-mean-square\_deviation} is adopted as measurement. The results is shown in \textbf{Table}~{\ref{tab: Model Calibration2}}.

 \begin{table}
  \caption{Model Calibration}
  \label{tab: Model Calibration2}
  \begin{tabular}{ccl}
    \toprule
    RMSE & Before & After\\
    \midrule
    CTR & 0.0957 & 0.0179\\
    Quit  & 0.6046 & 0.0077\\
  \bottomrule
\end{tabular}
\end{table}

\begin{figure}
\includegraphics[width=0.4\textwidth]{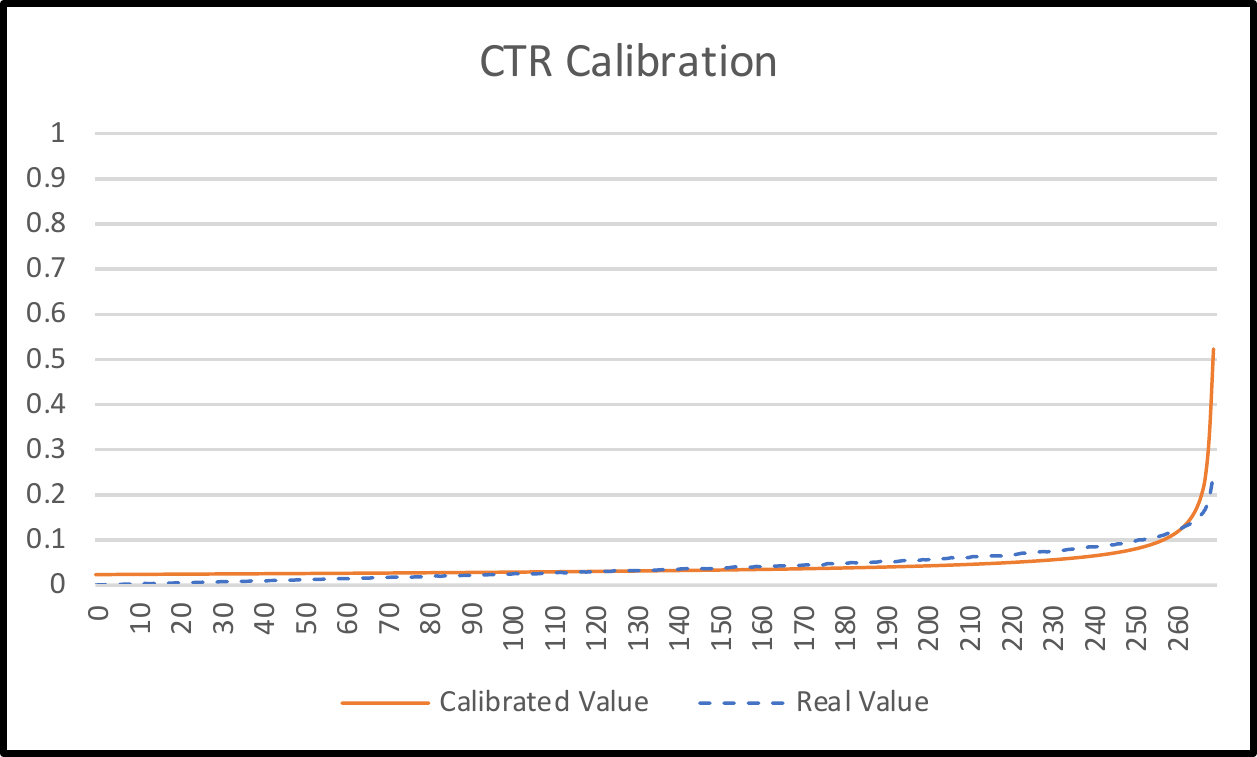}
\caption{Calibration of Click Model.}
\label{fig:ctrcali}
\end{figure}

\begin{figure}
\includegraphics[width=0.4\textwidth]{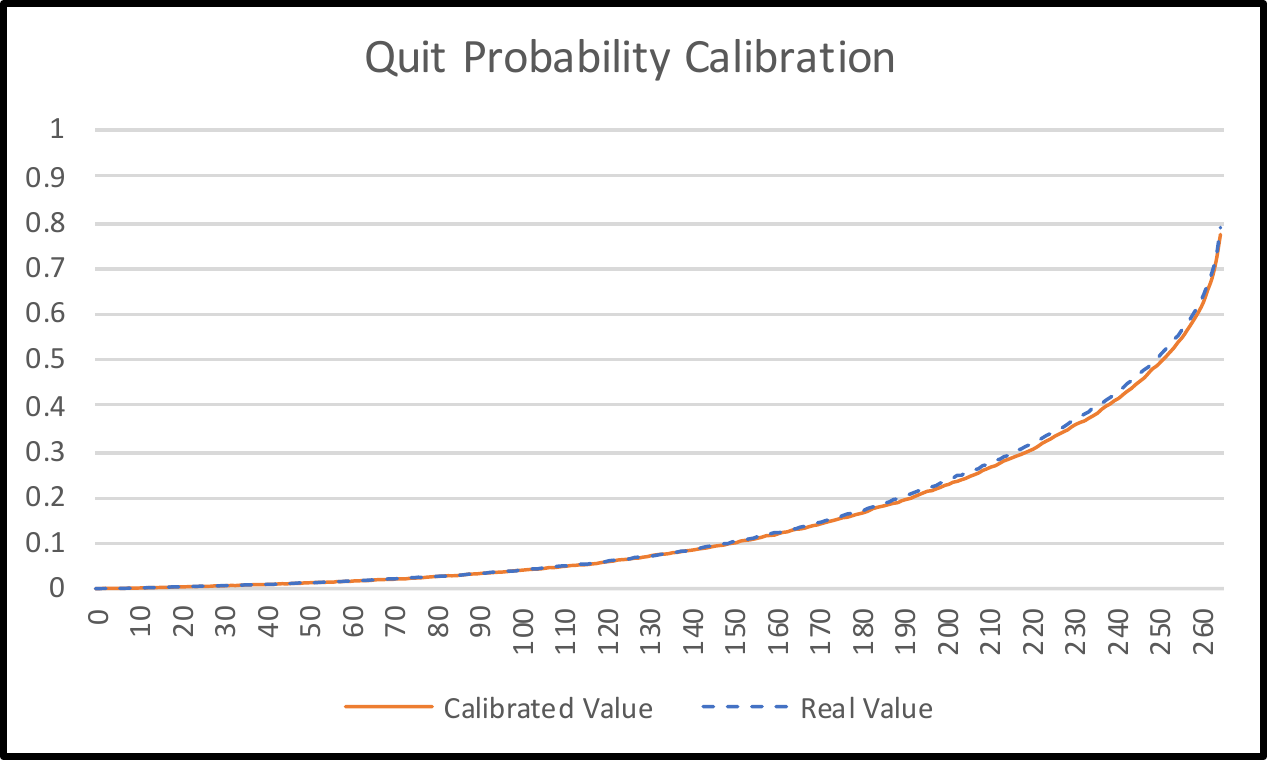}
\caption{Calibration of Quit Model}
\label{fig:quitcali}
\end{figure}

From \textbf{Table}~{\ref{tab: Model Calibration2}}, it can be seen significant improvement has been achieved after calibration, and the curve of real value and calibrated value is shown in \textbf{Figure}~{\ref{fig:ctrcali}} and \textbf{Figure}~{\ref{fig:quitcali}}. Abscissa axis is items sorted by the predicted scores, and vertical axis is calibrated score and the real score. The real score is obtained from items bin.

%
\subsubsection{Discrimination}

In Dataset 2, for each user we get the quit probability of corresponding candidates from the MDP Generator, i.e. the items in the user's browse session. Then a user's quit probability list $l_u=(q_1,...,q_i,...,q_n)$ is obtained, where $q_i$ is the quit probability when recommending $item\  i$ to $user\  u$. Standard Error (STD)
and MEAN
are calculated for each list, and the statistics of the dataset is shown in Table~\ref{tab: Discrimination of Browse Probabilites}. From the table, it can be demonstrated that, for each user, different candidates make different contributions to keep the user browsing, and they have a significant discrimination.
\begin{table}[h]
  \caption{Discrimination of Quit Probabilites}
  \label{tab: Discrimination of Browse Probabilites}
\centering
  \begin{tabular}{ccc}
    \toprule
    STD & MEAN & STD/MEAN\\
    \midrule
    0.1963 & 0.7348 & 0.3135 \\
  \bottomrule
\end{tabular}
\end{table}

\subsubsection{Weakly Related}

We further study the correlation between quit probability and immediate user engagement (i.e. each step reward). For each user, we get two item lists $l_{u1}$ and $l_{u2}$ with length $L=20$. $l_{u1}$ and $l_{u2}$ are formed greedily according to $R_{s_t, a_t}$ and $(1-P_{s_i, a_i, s_A})$ respectively. If $(1-P_{s_i, a_i, s_A})$ and $R_{s_t, a_t}$ are completely positive correlation, $l_{u1}$ and $l_{u2}$ will be the same, which leads to the equality of SSP and Greedy. We use Jaccard Index\footnote{https://en.wikipedia.org/wiki/Jaccard\_index} and NDCG\footnote{https://en.wikipedia.org/wiki/Discounted\_cumulative\_gain} to measure the similarity of $l_{u1}$ and $l_{u2}$, and the average result of the dataset is shown in Table~\ref{tab: Corrleation}. From the table, we find that in the dataset quit probability and immediate user engagement are weakly related.

\begin{table}[h]
  \caption{The Correlation between Quit Probability and CTR}
  \label{tab: Corrleation}
\centering
  \begin{tabular}{cccc}
    \toprule
    Mean Length & List Length & Jaccard Index & NDCG\\
    \midrule
    57 & 20 & 0.33 & 0.52 \\
  \bottomrule
\end{tabular}
\end{table}
\vspace{-0.5cm}

\subsection{SSP Planner: Offline Evaluation}

Now we deploy the strategies in Dataset 2.

\subsubsection{SSP Plan}
We plan a sequence list with $T$ steps: $L_{T}=$ $(a_1,a_2, \cdots, a_T)$, according to each strategy mentioned above. The revenue of $L_{T}$ can be calculated according to Equation~{(\ref{eqn:IPV3})}-{(\ref{eqn:ctr})}.

The detailed results are shown in Table~{\ref{tab: IPV of Offline Recommendation}}, and we can find that:
\begin {itemize}
\item Greedy achieves the best $CTR$, while SSP achieves the best $IPV$ and $BL$. This demonstrates our idea that $IPV$ can be improved through making users browse more. SSP does not aim to optimize each step effectiveness, and its purpose is to improve the total amount of cumulative clicks.
\item The longer the number of steps, the greater the advantages on $IPV$ and $BL$. See $T=20$ and $T=50$, when $T$ times 2.5, from 20 to 50, the improvement of both $IPV$ and $BL$ are more than 2.5 times($1347.57\  vs\  392.97$, $4045.47 \ vs\ 1066.08$ respectively). This result is in line with our expectation as planning more steps could lead to a bigger effect of the quit probability on users.

\end {itemize}

 \begin{table} [h]
  \caption{IPV of Offline Evaluation}\small
  \label{tab: IPV of Offline Recommendation}
\centering
\setlength{\tabcolsep}{1mm}{
  \begin{tabular}{ccccccc}
    \toprule
    \multirow{2}{*}{Method} & \multicolumn{3}{c}{T=20} & \multicolumn{3}{c}{T=50} \\
	& IPV & BL & CTR & IPV & BL & CTR\\
    \midrule
    GREEDY & 168.10 & 455.36  & {\bf 0.37} & 280.63  & 765.79 & {\bf 0.37}\\
    Beam Search & 321.91 & 977.78  & 0.33 & 660.90  & 2039.88 & 0.32 \\
    SSP & {\bf 392.97} & {\bf 1347.57}  & 0.29 & {\bf 1066.08}  & {\bf 4045.47} & 0.26 \\
  \bottomrule
\end{tabular}}
\end{table}
\vspace{-0.5cm}

\subsubsection{SSP Plan with Duplicate Removal}
In some practical scenarios, items are forbidden to display repeatedly. We need to make a compromise on the three strategies.
\begin{itemize}
\item Greedy: The items selected in the previous $t$ steps should be removed from the candidate set for step $t+1$.
\item Beam Search:  The items selected in the previous $t$ steps should be removed from the candidate set for step $t+1$.
\item SSP: When planing, we plan from step $T$ to step $1$ according to the upper bound of $V^*(s_t)$ of each step,
and keep the optimal $T$ items as the step's candidates at each step. 
When selecting, we do the selection from step $1$ to step $N$. Specifically, we choose the optimal one item and remove it from the remaining steps' candidates simultaneously.
\end{itemize}
From the detailed results in Table~{\ref{tab: IPV of Offline Recommendation with Duplicate Removal}}, we can find that although the compromises hurts the ideal effects, SSP still outperforms Greedy and Beam Search.

 \begin{table} [h]
  \caption{IPV of Offline Evaluation with Deduplication}\small
  \label{tab: IPV of Offline Recommendation with Duplicate Removal}
\centering
\setlength{\tabcolsep}{1mm}{
  \begin{tabular}{ccccccc}
    \toprule
    \multirow{2}{*}{Method} & \multicolumn{3}{c}{T=20} & \multicolumn{3}{c}{T=50} \\
	& IPV & BL & CTR & IPV & BL & CTR\\
    \midrule
    GREEDY & 68.06 & 216.38  & {\bf 0.31} & 68.23  & 217.93 & {\bf 0.31}  \\
    Beam Search & 105.52 & 427.80  & 0.25 & 107.06  & 439.59 & 0.24 \\
    SSP & {\bf 189.11} & {\bf 999.51}  & 0.19 & {\bf 242.77}  & {\bf 1632.56} & 0.15\\
  \bottomrule
\end{tabular}}
\end{table}
\vspace{-0.4cm}

\subsubsection{SSP Plan with Noise}
Since there may exist a gap between offline environment and online environment, which makes the predicted click-through rate and quit probability offline are not absolutely equivalent to the real value online, we introduce a set of noise experiments before deploying {\it MDP-SSP} online.

\begin{figure}[h]
\setlength{\abovecaptionskip}{0cm}   
\setlength{\belowcaptionskip}{-0.5cm}   
\centering
\subfigure[T=20]{\includegraphics[width=0.2\textwidth]{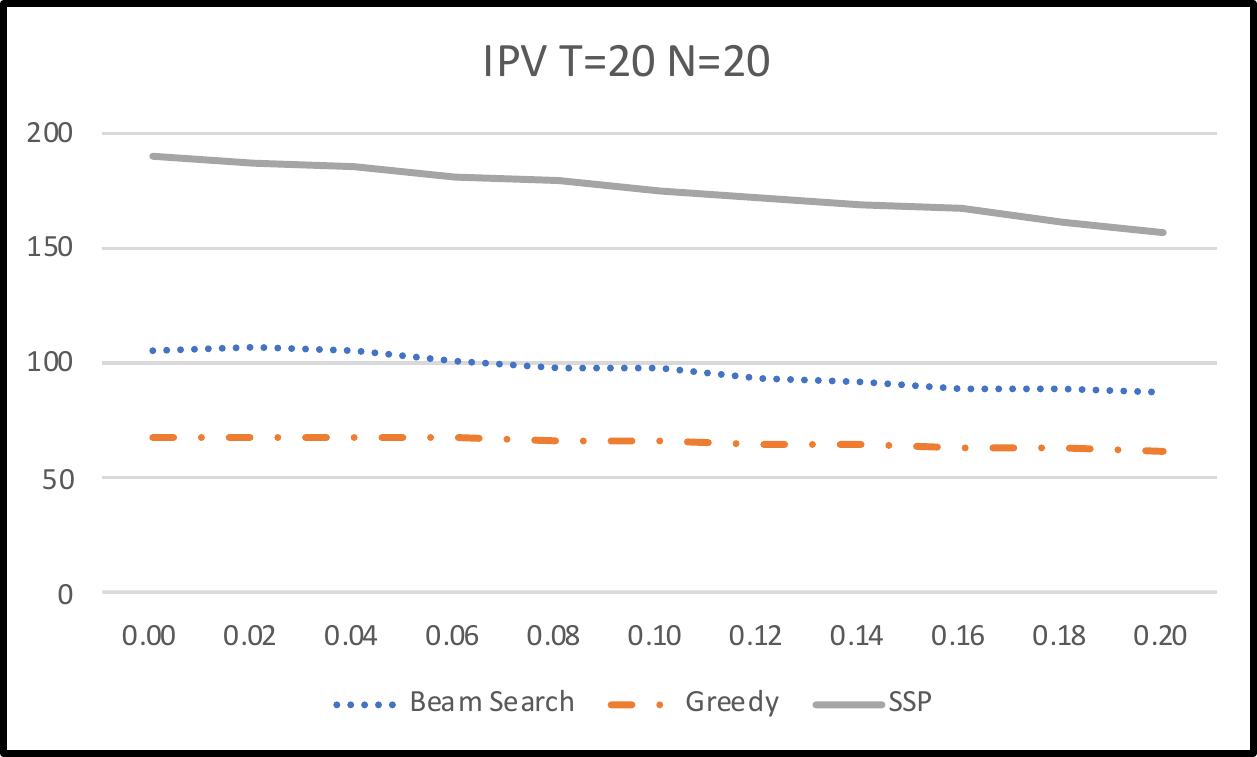}}
\subfigure[T=50]{\includegraphics[width=0.2\textwidth]{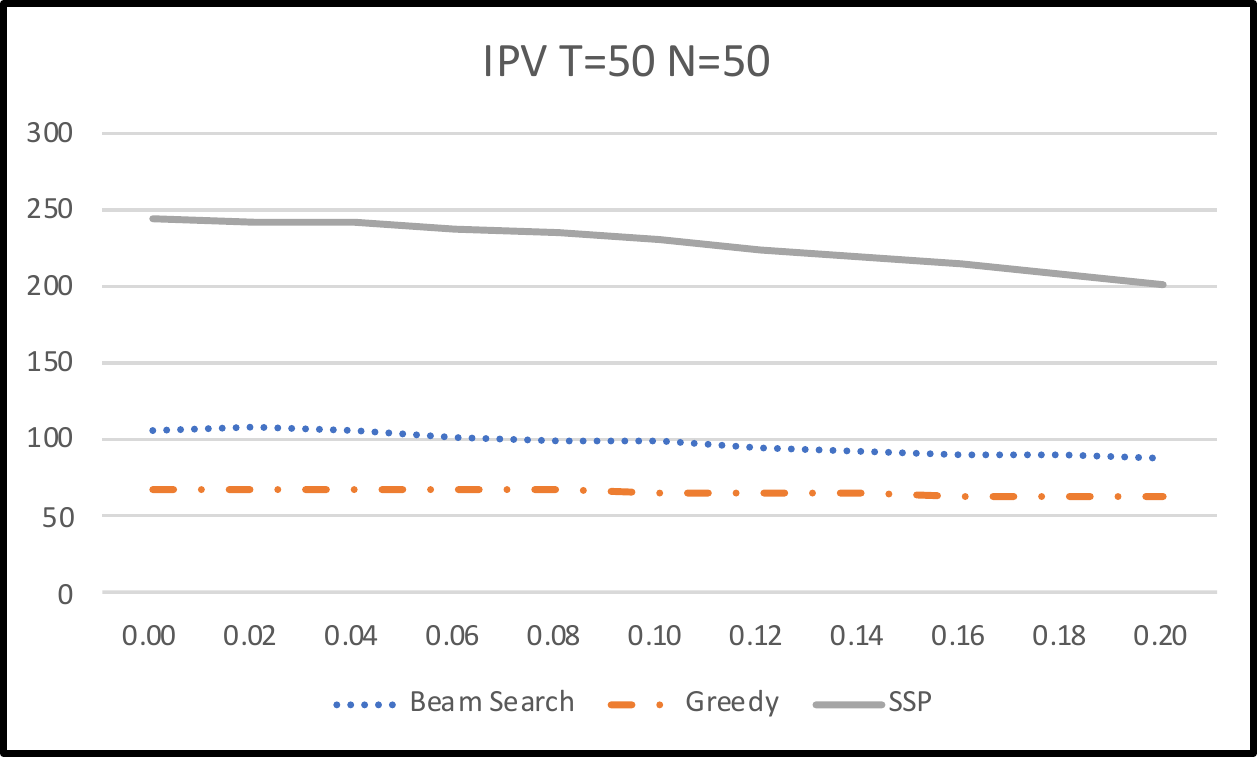}}
\caption{Noise Experiments}
   \label{fig:noise}
\end{figure}

The experiments are conducted in the following way: we add random noises in the click-through rate and quit probability given by the offline environment. Assuming the noise $e \sim U(a, b)$ where $U(a, b)$ is a uniform distribution, we define $a=-0.02m$, $b=0.02m$, where $m$ is an integer ranges from $0$ to $10$. We plan according to the value with noise, and calculate the final revenue with the real value. The results are shown in Figure~{\ref{fig:noise}}. The horizontal axis represents the noise, i.e. $b$ in $U(a, b)$, and the vertical axis is the revenue, i.e. cumulative clicks.

From Figure~{\ref{fig:noise}} we can find that although SSP is more sensitive to noise, it performs better than Greedy and Beam Search. It demonstrates that considering quit probability plays a very important role in {\it IPV} issue.

\subsection{SSP Planner: Online Evaluation}

For online evaluation, we deployed SSP and Greedy strategies on a real E-commerce APP. For further comparison, we conduct a experiments with quit model which does not introduce MIL, and the strategy is named  $SSP_{no\_{MIL}}$. Three strategies run online with the same traffic for one week, and the results shown in Table~{\ref{tab: IPV of online Recommendation}}\footnote{The data has been processed for business reason.} demonstrate that:
\begin{itemize}
\item For cumulative clicks, quit probability cannot be ignored in sequential recommendations, see SSP and Greedy.
\item The accuracy of quit probability directly influence the results, see SSP and $SSP_{no\_{MIL}}$.
\end{itemize}

 \begin{table} [h]
  \caption{IPV of online Evaluation}
  \label{tab: IPV of online Recommendation}
\centering
  \begin{tabular}{ccc}
    \toprule
    Method & IPV & BL\\
    \midrule
    Greedy & 0.9296 & 0.9440 \\
    $SSP_{no\_MIL}$ & 0.9638 & 0.9789 \\
    SSP & 1 & 1 \\
  \bottomrule
\end{tabular}
\end{table}

\section{Conclusions}
In this paper, we study the problem of maximizing cumulative user engagement in sequential recommendation where the browse length is not fixed. Furthermore, we propose an online optimization framework, in which the problem can be reduced to a SSP problem. Then a practical approach that is easy to implement in real-world applications is proposed, and the corresponding optimization problem can be efficiently solved via dynamic programming. The superior advantage of our method is also verified with both offline and online experiments by generating optimal personalized recommendations.
%

In the future, we will study the {\it MDP-SSP} with deduplication. While the current {\it MDP-SSP} could yield good sequential recommendation, it fails to consider the item duplication issue, which is commonly not allowed in practice. Although we propose a compromise strategy and it outperforms Beam Search and Greedy strategy (which are commonly used in practice), it is not the optimal solution when considering items deduplication.
It will bring more insights if the deduplication constraint can be modeled into the strategy.

\clearpage


\begin{thebibliography}{26}


\ifx \showCODEN    \undefined \def \showCODEN     #1{\unskip}     \fi
\ifx \showDOI      \undefined \def \showDOI       #1{#1}\fi
\ifx \showISBNx    \undefined \def \showISBNx     #1{\unskip}     \fi
\ifx \showISBNxiii \undefined \def \showISBNxiii  #1{\unskip}     \fi
\ifx \showISSN     \undefined \def \showISSN      #1{\unskip}     \fi
\ifx \showLCCN     \undefined \def \showLCCN      #1{\unskip}     \fi
\ifx \shownote     \undefined \def \shownote      #1{#1}          \fi
\ifx \showarticletitle \undefined \def \showarticletitle #1{#1}   \fi
\ifx \showURL      \undefined \def \showURL       {\relax}        \fi
\providecommand\bibfield[2]{#2}
\providecommand\bibinfo[2]{#2}
\providecommand\natexlab[1]{#1}
\providecommand\showeprint[2][]{arXiv:#2}

\bibitem[\protect\citeauthoryear{Amores}{Amores}{2013}]%
        {amores2013multiple}
\bibfield{author}{\bibinfo{person}{Jaume Amores}.}
  \bibinfo{year}{2013}\natexlab{}.
\newblock \showarticletitle{Multiple instance classification: review, taxonomy
  and comparative study}.
\newblock \bibinfo{journal}{\emph{Artificial intelligence}}
  \bibinfo{volume}{201} (\bibinfo{year}{2013}), \bibinfo{pages}{81--105}.
\newblock


\bibitem[\protect\citeauthoryear{Andrews, Tsochantaridis, and Hofmann}{Andrews
  et~al\mbox{.}}{2003}]%
        {andrews2003support}
\bibfield{author}{\bibinfo{person}{Stuart Andrews}, \bibinfo{person}{Ioannis
  Tsochantaridis}, {and} \bibinfo{person}{Thomas Hofmann}.}
  \bibinfo{year}{2003}\natexlab{}.
\newblock \showarticletitle{Support vector machines for multiple-instance
  learning}. In \bibinfo{booktitle}{\emph{Advances in Neural Information
  Processing Systems}}. \bibinfo{address}{Vancouver, Canada},
  \bibinfo{pages}{577--584}.
\newblock


\bibitem[\protect\citeauthoryear{Barto, Bradtke, and Singh}{Barto
  et~al\mbox{.}}{1995}]%
        {barto1995learning}
\bibfield{author}{\bibinfo{person}{Andrew~G Barto}, \bibinfo{person}{Steven~J
  Bradtke}, {and} \bibinfo{person}{Satinder~P Singh}.}
  \bibinfo{year}{1995}\natexlab{}.
\newblock \showarticletitle{Learning to act using real-time dynamic
  programming}.
\newblock \bibinfo{journal}{\emph{Artificial Intelligence}}
  \bibinfo{volume}{72}, \bibinfo{number}{1-2} (\bibinfo{year}{1995}),
  \bibinfo{pages}{81--138}.
\newblock


\bibitem[\protect\citeauthoryear{Bertsekas and Tsitsiklis}{Bertsekas and
  Tsitsiklis}{1991}]%
        {bertsekas1991analysis}
\bibfield{author}{\bibinfo{person}{Dimitri~P Bertsekas} {and}
  \bibinfo{person}{John~N Tsitsiklis}.} \bibinfo{year}{1991}\natexlab{}.
\newblock \showarticletitle{An analysis of stochastic shortest path problems}.
\newblock \bibinfo{journal}{\emph{Mathematics of Operations Research}}
  \bibinfo{volume}{16}, \bibinfo{number}{3} (\bibinfo{year}{1991}),
  \bibinfo{pages}{580--595}.
\newblock


\bibitem[\protect\citeauthoryear{Bonet and Geffner}{Bonet and Geffner}{2003}]%
        {bonet2003labeled}
\bibfield{author}{\bibinfo{person}{Blai Bonet} {and} \bibinfo{person}{Hector
  Geffner}.} \bibinfo{year}{2003}\natexlab{}.
\newblock \showarticletitle{Labeled RTDP: improving the convergence of
  real-time dynamic programming.}. In \bibinfo{booktitle}{\emph{Proceedings of
  the Thirteenth International Conference on Automated Planning and
  Scheduling}}, Vol.~\bibinfo{volume}{3}. \bibinfo{address}{Trento, Italy},
  \bibinfo{pages}{12--21}.
\newblock


\bibitem[\protect\citeauthoryear{Bunescu and Mooney}{Bunescu and
  Mooney}{2007}]%
        {bunescu2007multiple}
\bibfield{author}{\bibinfo{person}{Razvan~C Bunescu} {and}
  \bibinfo{person}{Raymond~J Mooney}.} \bibinfo{year}{2007}\natexlab{}.
\newblock \showarticletitle{Multiple instance learning for sparse positive
  bags}. In \bibinfo{booktitle}{\emph{Proceedings of the Twenty-Fourth
  International Conference on Machine Learning}}. ACM,
  \bibinfo{address}{Corvallis, Oregon, USA}, \bibinfo{pages}{105--112}.
\newblock


\bibitem[\protect\citeauthoryear{Carbonell and Goldstein}{Carbonell and
  Goldstein}{1998}]%
        {carbonell1998use}
\bibfield{author}{\bibinfo{person}{Jaime Carbonell} {and} \bibinfo{person}{Jade
  Goldstein}.} \bibinfo{year}{1998}\natexlab{}.
\newblock \showarticletitle{The use of MMR, diversity-based reranking for
  reordering documents and producing summaries}. In
  \bibinfo{booktitle}{\emph{Proceedings of the Twentiy-First Annual
  International ACM SIGIR Conference on Research and Development in Information
  Retrieval}}. ACM, \bibinfo{address}{Melbourne, Australia},
  \bibinfo{pages}{335--336}.
\newblock


\bibitem[\protect\citeauthoryear{Chen, Xu, Zhang, Tang, Cao, Qin, and Zha}{Chen
  et~al\mbox{.}}{2018}]%
        {chen2018sequential}
\bibfield{author}{\bibinfo{person}{Xu Chen}, \bibinfo{person}{Hongteng Xu},
  \bibinfo{person}{Yongfeng Zhang}, \bibinfo{person}{Jiaxi Tang},
  \bibinfo{person}{Yixin Cao}, \bibinfo{person}{Zheng Qin}, {and}
  \bibinfo{person}{Hongyuan Zha}.} \bibinfo{year}{2018}\natexlab{}.
\newblock \showarticletitle{Sequential recommendation with user memory
  networks}. In \bibinfo{booktitle}{\emph{Proceedings of the Eleventh ACM
  International Conference on Web Search and Data Mining}}. ACM,
  \bibinfo{address}{Los Angeles, California, USA}, \bibinfo{pages}{108--116}.
\newblock


\bibitem[\protect\citeauthoryear{Devooght and Bersini}{Devooght and
  Bersini}{2017}]%
        {devooght2017long}
\bibfield{author}{\bibinfo{person}{Robin Devooght} {and}
  \bibinfo{person}{Hugues Bersini}.} \bibinfo{year}{2017}\natexlab{}.
\newblock \showarticletitle{Long and short-term recommendations with recurrent
  neural networks}. In \bibinfo{booktitle}{\emph{Proceedings of The
  Twenty-Fifth Conference on User Modeling, Adaptation and Personalization}}.
  ACM, \bibinfo{address}{Bratislava, Slovakia}, \bibinfo{pages}{13--21}.
\newblock


\bibitem[\protect\citeauthoryear{Dietterich, Lathrop, and
  Lozano-P{\'e}rez}{Dietterich et~al\mbox{.}}{1997}]%
        {dietterich1997solving}
\bibfield{author}{\bibinfo{person}{Thomas~G Dietterich},
  \bibinfo{person}{Richard~H Lathrop}, {and} \bibinfo{person}{Tom{\'a}s
  Lozano-P{\'e}rez}.} \bibinfo{year}{1997}\natexlab{}.
\newblock \showarticletitle{Solving the multiple instance problem with
  axis-parallel rectangles}.
\newblock \bibinfo{journal}{\emph{Artificial Intelligence}}
  \bibinfo{volume}{89}, \bibinfo{number}{1-2} (\bibinfo{year}{1997}),
  \bibinfo{pages}{31--71}.
\newblock


\bibitem[\protect\citeauthoryear{Donkers, Loepp, and Ziegler}{Donkers
  et~al\mbox{.}}{2017}]%
        {donkers2017sequential}
\bibfield{author}{\bibinfo{person}{Tim Donkers}, \bibinfo{person}{Benedikt
  Loepp}, {and} \bibinfo{person}{J{\"u}rgen Ziegler}.}
  \bibinfo{year}{2017}\natexlab{}.
\newblock \showarticletitle{Sequential user-based recurrent neural network
  recommendations}. In \bibinfo{booktitle}{\emph{Proceedings of the Eleventh
  ACM Conference on Recommender Systems}}. ACM, \bibinfo{address}{Como, Italy},
  \bibinfo{pages}{152--160}.
\newblock


\bibitem[\protect\citeauthoryear{Ebesu, Shen, and Fang}{Ebesu
  et~al\mbox{.}}{2018}]%
        {ebesu2018collaborative}
\bibfield{author}{\bibinfo{person}{Travis Ebesu}, \bibinfo{person}{Bin Shen},
  {and} \bibinfo{person}{Yi Fang}.} \bibinfo{year}{2018}\natexlab{}.
\newblock \showarticletitle{Collaborative memory network for recommendation
  systems}.
\newblock \bibinfo{journal}{\emph{arXiv preprint arXiv:1804.10862}}
  (\bibinfo{year}{2018}).
\newblock


\bibitem[\protect\citeauthoryear{G{\"a}rtner, Flach, Kowalczyk, and
  Smola}{G{\"a}rtner et~al\mbox{.}}{2002}]%
        {gartner2002multi}
\bibfield{author}{\bibinfo{person}{Thomas G{\"a}rtner},
  \bibinfo{person}{Peter~A Flach}, \bibinfo{person}{Adam Kowalczyk}, {and}
  \bibinfo{person}{Alexander~J Smola}.} \bibinfo{year}{2002}\natexlab{}.
\newblock \showarticletitle{Multi-instance kernels}. In
  \bibinfo{booktitle}{\emph{Proceedings of the Nineteenth International
  Conference on Machine Learning}}, Vol.~\bibinfo{volume}{2}.
  \bibinfo{address}{Sydney, Australia}, \bibinfo{pages}{179--186}.
\newblock


\bibitem[\protect\citeauthoryear{He, Pan, Jin, Xu, Liu, Xu, Shi, Atallah,
  Herbrich, Bowers, et~al\mbox{.}}{He et~al\mbox{.}}{2014}]%
        {he2014practical}
\bibfield{author}{\bibinfo{person}{Xinran He}, \bibinfo{person}{Junfeng Pan},
  \bibinfo{person}{Ou Jin}, \bibinfo{person}{Tianbing Xu}, \bibinfo{person}{Bo
  Liu}, \bibinfo{person}{Tao Xu}, \bibinfo{person}{Yanxin Shi},
  \bibinfo{person}{Antoine Atallah}, \bibinfo{person}{Ralf Herbrich},
  \bibinfo{person}{Stuart Bowers}, {et~al\mbox{.}}}
  \bibinfo{year}{2014}\natexlab{}.
\newblock \showarticletitle{Practical lessons from predicting clicks on ads at
  facebook}. In \bibinfo{booktitle}{\emph{Proceedings of the Eighth
  International Workshop on Data Mining for Online Advertising}}. ACM,
  \bibinfo{address}{New York, NY, USA}, \bibinfo{pages}{1--9}.
\newblock


\bibitem[\protect\citeauthoryear{Hidasi, Karatzoglou, Baltrunas, and
  Tikk}{Hidasi et~al\mbox{.}}{2015}]%
        {hidasi2015session}
\bibfield{author}{\bibinfo{person}{Bal{\'a}zs Hidasi},
  \bibinfo{person}{Alexandros Karatzoglou}, \bibinfo{person}{Linas Baltrunas},
  {and} \bibinfo{person}{Domonkos Tikk}.} \bibinfo{year}{2015}\natexlab{}.
\newblock \showarticletitle{Session-based recommendations with recurrent neural
  networks}.
\newblock \bibinfo{journal}{\emph{arXiv preprint arXiv:1511.06939}}
  (\bibinfo{year}{2015}).
\newblock


\bibitem[\protect\citeauthoryear{Huang, Zhao, Dou, Wen, and Chang}{Huang
  et~al\mbox{.}}{2018}]%
        {huang2018improving}
\bibfield{author}{\bibinfo{person}{Jin Huang}, \bibinfo{person}{Wayne~Xin
  Zhao}, \bibinfo{person}{Hongjian Dou}, \bibinfo{person}{Ji-Rong Wen}, {and}
  \bibinfo{person}{Edward~Y Chang}.} \bibinfo{year}{2018}\natexlab{}.
\newblock \showarticletitle{Improving sequential recommendation with
  knowledge-enhanced memory networks}. In \bibinfo{booktitle}{\emph{Proceedings
  of the Forty-First International ACM SIGIR Conference on Research \&
  Development in Information Retrieval}}. ACM, \bibinfo{address}{Ann Arbor
  Michigan, USA}, \bibinfo{pages}{505--514}.
\newblock


\bibitem[\protect\citeauthoryear{Kolobov, Mausam, Weld, and Geffner}{Kolobov
  et~al\mbox{.}}{2011}]%
        {kolobov2011heuristic}
\bibfield{author}{\bibinfo{person}{Andrey Kolobov}, \bibinfo{person}{Mausam
  Mausam}, \bibinfo{person}{Daniel~S Weld}, {and} \bibinfo{person}{Hector
  Geffner}.} \bibinfo{year}{2011}\natexlab{}.
\newblock \showarticletitle{Heuristic search for generalized stochastic
  shortest path MDPs}. In \bibinfo{booktitle}{\emph{Proceedings of the
  Twenty-First International Conference on Automated Planning and Scheduling}}.
  \bibinfo{address}{Freiburg, Germany}.
\newblock


\bibitem[\protect\citeauthoryear{Lin, Lin, and Weng}{Lin et~al\mbox{.}}{2007}]%
        {lin2007note}
\bibfield{author}{\bibinfo{person}{Hsuan-Tien Lin}, \bibinfo{person}{Chih-Jen
  Lin}, {and} \bibinfo{person}{Ruby~C Weng}.} \bibinfo{year}{2007}\natexlab{}.
\newblock \showarticletitle{A note on Platt’s probabilistic outputs for
  support vector machines}.
\newblock \bibinfo{journal}{\emph{Machine learning}} \bibinfo{volume}{68},
  \bibinfo{number}{3} (\bibinfo{year}{2007}), \bibinfo{pages}{267--276}.
\newblock


\bibitem[\protect\citeauthoryear{Maron and Lozano-P{\'e}rez}{Maron and
  Lozano-P{\'e}rez}{1998}]%
        {maron1998framework}
\bibfield{author}{\bibinfo{person}{Oded Maron} {and} \bibinfo{person}{Tom{\'a}s
  Lozano-P{\'e}rez}.} \bibinfo{year}{1998}\natexlab{}.
\newblock \showarticletitle{A framework for multiple-instance learning}. In
  \bibinfo{booktitle}{\emph{Advances in Neural Information Processing
  Systems}}. \bibinfo{address}{Massachusetts, USA}, \bibinfo{pages}{570--576}.
\newblock


\bibitem[\protect\citeauthoryear{Niculescu-Mizil and Caruana}{Niculescu-Mizil
  and Caruana}{2005}]%
        {niculescu2005predicting}
\bibfield{author}{\bibinfo{person}{Alexandru Niculescu-Mizil} {and}
  \bibinfo{person}{Rich Caruana}.} \bibinfo{year}{2005}\natexlab{}.
\newblock \showarticletitle{Predicting good probabilities with supervised
  learning}. In \bibinfo{booktitle}{\emph{Proceedings of the Twenty-Second
  International Conference on Machine Learning}}. ACM, \bibinfo{address}{Bonn,
  Germany}, \bibinfo{pages}{625--632}.
\newblock


\bibitem[\protect\citeauthoryear{Polychronopoulos and
  Tsitsiklis}{Polychronopoulos and Tsitsiklis}{1996}]%
        {polychronopoulos1996stochastic}
\bibfield{author}{\bibinfo{person}{George~H Polychronopoulos} {and}
  \bibinfo{person}{John~N Tsitsiklis}.} \bibinfo{year}{1996}\natexlab{}.
\newblock \showarticletitle{Stochastic shortest path problems with recourse}.
\newblock \bibinfo{journal}{\emph{Networks: An International Journal}}
  \bibinfo{volume}{27}, \bibinfo{number}{2} (\bibinfo{year}{1996}),
  \bibinfo{pages}{133--143}.
\newblock


\bibitem[\protect\citeauthoryear{Tang and Wang}{Tang and Wang}{2018}]%
        {tang2018personalized}
\bibfield{author}{\bibinfo{person}{Jiaxi Tang} {and} \bibinfo{person}{Ke
  Wang}.} \bibinfo{year}{2018}\natexlab{}.
\newblock \showarticletitle{Personalized top-n sequential recommendation via
  convolutional sequence embedding}. In \bibinfo{booktitle}{\emph{Proceedings
  of the Eleventh ACM International Conference on Web Search and Data Mining}}.
  ACM, \bibinfo{address}{Los Angeles, California, USA},
  \bibinfo{pages}{565--573}.
\newblock


\bibitem[\protect\citeauthoryear{Teo, Nassif, Hill, Srinivasan, Goodman, Mohan,
  and Vishwanathan}{Teo et~al\mbox{.}}{2016}]%
        {teo2016adaptive}
\bibfield{author}{\bibinfo{person}{Choon~Hui Teo}, \bibinfo{person}{Houssam
  Nassif}, \bibinfo{person}{Daniel Hill}, \bibinfo{person}{Sriram Srinivasan},
  \bibinfo{person}{Mitchell Goodman}, \bibinfo{person}{Vijai Mohan}, {and}
  \bibinfo{person}{SVN Vishwanathan}.} \bibinfo{year}{2016}\natexlab{}.
\newblock \showarticletitle{Adaptive, personalized diversity for visual
  discovery}. In \bibinfo{booktitle}{\emph{Proceedings of The Tenth ACM
  Conference on Recommender Systems}}. ACM, \bibinfo{address}{Boston, MA, USA},
  \bibinfo{pages}{35--38}.
\newblock


\bibitem[\protect\citeauthoryear{Trevizan, Thi{\'e}baux, Santana, and
  Williams}{Trevizan et~al\mbox{.}}{2016}]%
        {trevizan2016heuristic}
\bibfield{author}{\bibinfo{person}{Felipe~W Trevizan}, \bibinfo{person}{Sylvie
  Thi{\'e}baux}, \bibinfo{person}{Pedro~Henrique Santana}, {and}
  \bibinfo{person}{Brian~Charles Williams}.} \bibinfo{year}{2016}\natexlab{}.
\newblock \showarticletitle{Heuristic search in dual space for constrained
  stochastic shortest path problems.}. In \bibinfo{booktitle}{\emph{Proceedings
  of the Thirteenth International Conference on Automated Planning and
  Scheduling}}. \bibinfo{address}{London, UK}, \bibinfo{pages}{326--334}.
\newblock


\bibitem[\protect\citeauthoryear{Wu, Ahmed, Beutel, Smola, and Jing}{Wu
  et~al\mbox{.}}{2017}]%
        {wu2017recurrent}
\bibfield{author}{\bibinfo{person}{Chao-Yuan Wu}, \bibinfo{person}{Amr Ahmed},
  \bibinfo{person}{Alex Beutel}, \bibinfo{person}{Alexander~J Smola}, {and}
  \bibinfo{person}{How Jing}.} \bibinfo{year}{2017}\natexlab{}.
\newblock \showarticletitle{Recurrent recommender networks}. In
  \bibinfo{booktitle}{\emph{Proceedings of The Tenth ACM International
  Conference on Web Search and Data Mining}}. ACM, \bibinfo{address}{Cambridge,
  UK}, \bibinfo{pages}{495--503}.
\newblock


\bibitem[\protect\citeauthoryear{Zhang and Goldman}{Zhang and Goldman}{2002}]%
        {zhang2002dd}
\bibfield{author}{\bibinfo{person}{Qi Zhang} {and} \bibinfo{person}{Sally~A
  Goldman}.} \bibinfo{year}{2002}\natexlab{}.
\newblock \showarticletitle{EM-DD: An improved multiple-instance learning
  technique}. In \bibinfo{booktitle}{\emph{Advances in Neural Information
  Processing Systems}}. \bibinfo{address}{Vancouver, British Columbia, Canada},
  \bibinfo{pages}{1073--1080}.
\newblock


\end{thebibliography}


\appendix

\end{document}